\documentclass[a4paper]{jpconf}
\usepackage{graphicx}
\usepackage{hyperref}
\usepackage{url}
\usepackage{epstopdf}
\begin{document}
\title{Modeling of two-particle femtoscopic correlations at top RHIC energy}

\author{N Ermakov$^1$ and G Nigmatkulov$^1$}

\address{$^1$ National Research Nuclear University MEPhI (Moscow
  Engineering Physics Institute), Kashirskoe highway 31, Moscow, 115409, Russia}
\ead{noermakov@mephi.ru, ganigmatkulov@mephi.ru}

\begin{abstract}
The spatial and temporal characteristics of particle emitting source
produced in particle and/or nuclear collisions can be measured by
using two-particle 
femtoscopic correlations. These correlations arise due to quantum
statistics, Coulomb and strong final state interactions.
In this paper we report on the calculations of like-sign pion
femtoscopic correlations produced in p+p, p+Au, d+Au, Au+Au at top
RHIC energy using Ultra Relativistic Quantum Molecular Dynamics Model
(UrQMD). Three-dimensional correlation functions are constructed using
the Bertsch-Pratt parametrization of the two-particle relative
momentum. The correlation functions are studied in several transverse
mass ranges. The emitting source radii of charged pions, $R_{out}$,
$R_{side}$, $R_{long}$, are obtained from Gaussian fit
to the correlation functions and compared to data from the STAR and
PHENIX experiments.
\end{abstract}

\section{Introduction}

The spatiotemporal structures of particle emitting source in
high-energy collisions are essentially defined by the dynamics
of the collision processes~\cite{lcms_0}. The femtoscopy method
allows to measure the  spatial and temporal characteristics of
emitting region in high-energy collisions.
Such correlations arise due to quantum statistics, Coulomb and
strong final state interactions.

We present calculations of the femtoscopic radii of two-pion
correlations (referred to as Bose-Einstein, or Hanbury-Brown Twiss
``HBT'', correlations) in p+p and central (0-10\% centrality) p+Au,
d+Au, Au+Au collisions at top RHIC energy $\sqrt{s_{NN}} = 200$ GeV.
the Ultra Relativistic Quantum Molecular
Dynamics Model (UrQMD)~\cite{urqmd_0,urqmd_1,urqmd_2,urqmd_3} was
used to simulate the ion collisions. The calculated femtoscopic radii
for Au+Au collisions were compared to the STAR~\cite{star_exp} and PHENIX experimental
data~\cite{phenix_exp}.

\section{Femtoscopy}

The method of femtoscopy is created to measure the space-time extents
of the particle emitting region at freeze-out. It is based on
measurements of identical particle correlations (usually pions). The
femtoscopic correlations are calculated as a function of relative
momentum, expressed as $\mathbf{q} = \mathbf{p_{1}} - \mathbf{p_{2}}$.
In order to estimate the particle emitting source parameters, one
uses the correlation function, $C(\mathbf{q})$. It is constracted as
follows: 
\begin{equation}
  C(\mathbf{q}) = \frac{A(\mathbf{q})}{B(\mathbf{q})} ,
  \label{eq2}
\end{equation}
where $A(\mathbf{q})$ is a distribution of two-particle relative
momentum that contains quantum statistical correlations, and
$B(\mathbf{q})$ is the reference distribution that has all
experimental effects (event, single-particle and pair-particle
selection criteria) as the first one except for the absence of the
Bose-Einstein correlations. 
Following the modern analysis techniques, we decompose the relative
momentum of the pairs into the three projections, namely $out$, $side$
and $long$, according to the Bettsch-Pratt
parameterization~\cite{osl1,osl2}. Correlation functions were constructed in
longitudinaly co-moving system (LCMS), where
$p_{z1}+p_{z2}=0$. Asuming a Gaussian emission profile of the source
the correlations functions are fitted with the form~\cite{lcms_0,lcms_1}:

\begin{equation}
  C(\mathbf{q}, \mathbf{k_{T}}) = 1 + \lambda(\mathbf{k_{T}})\exp\left[ - \sum_{i,j=out,side,long}
    q_{i}q_{j}R^{2}_{ij}(\mathbf{k_{T}}) \right] ,
  \label{eq1}
\end{equation}
where $\lambda$ is the fraction of correlated pairs and
$q_{i}$  is the elative momentum of the pair in the $i$ direction. The
longitudinal direction along the beam axis corresponds to the $long$
term, the outward direction is pointing along the transverse component
of the average momentum  $\mathbf{k}$ of a pair ($k_{T} =
|\mathbf{p_{1T}} + \mathbf{p_{2T}}|/2$) and the sideward direction
is orthogonal to both $out$ and $long$.
The effect of cross terms with $i \neq j$ in the HBT radii is
negligible as the pseudorapidity cuts $|\eta| < 1$ are used.
The HBT radii are related to regions of homogeneity. According to
Ref.~\cite{radii}, $R_{side}$ contains information about geometry,
$R_{out}$ convolutes the information about geometry and emission duration and
$R_{long}$ contains information about system lifetime.
The $m_{T}$ ($m_{T} = \sqrt{m^{2}_{\pi^{\pm}} + k_{T}^{2}}$) dependence
of the femtoscopic radii shows the dynamic of the system and allows to
probe the different regions of the homogeneity.

\section{Results and Discussions}

We present femtoscopic radii calculations of identical-pion
correlations in p+p and central (0-10\% centrality) p+Au, d+Au,
Au+Au collisions at top RHIC energy $\sqrt{s_{NN}} = 200$~GeV using
the UrQMD model. The collisions centrality was defined by comparing
impact parameter and multiplicity distributions of charged particles
within the typical acceptance of the collider experiments.

UrQMD is a microscopic many body approach and can be applied to study
hadron-hadron, hadron-nucleus and heavy ion reactions, at relativistic
energies. The UrQMD does not contain quantum statistics
correlations. In order to add these correlations, the simulated phase
space points of charged pions at their freeze-out time were passed
to a Correlation After Burner (CRAB) analyzing program~\cite{crab}. 

\begin{figure}[h]
  \centering
  \includegraphics[width=0.8\textwidth] {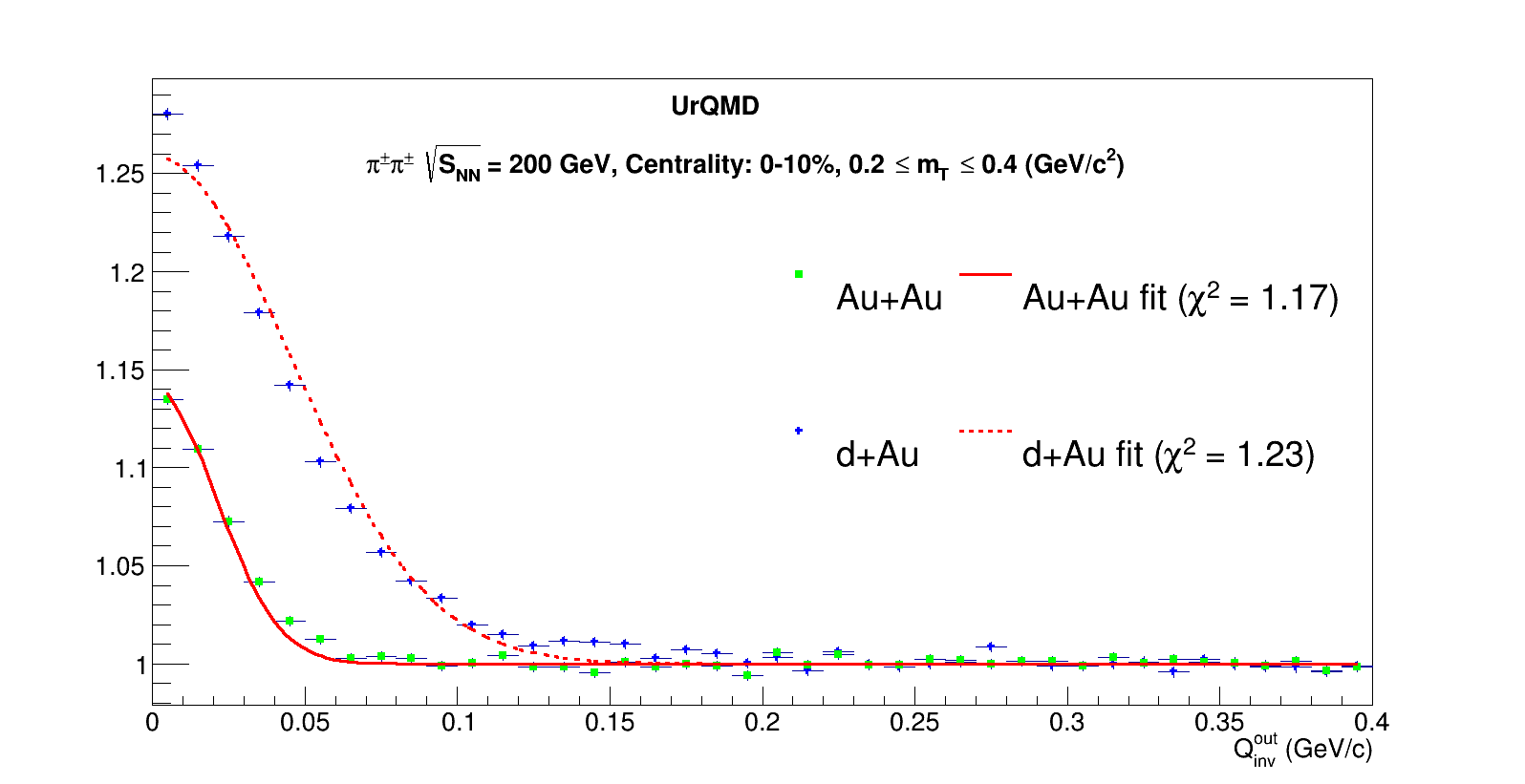}
  \caption{Example of the correlation function and projection of the
    three-dimensional Gaussian fit to it in the outward direction.}
  \label{fig1}
\end{figure}

Fig.~\ref{fig1} shows the example of the projection in the outward
direction of the correlation functions of charged pions calculated for
d+Au (blue rombs) and Au+Au (green squares) collisions in the
transverse mass range ($m_{T} =0.2-0.4$~GeV/{\it c}$^{2}$) and
corresponding to them fits.

\begin{figure}[h]
  \centering
  \includegraphics[width=0.8\textwidth] {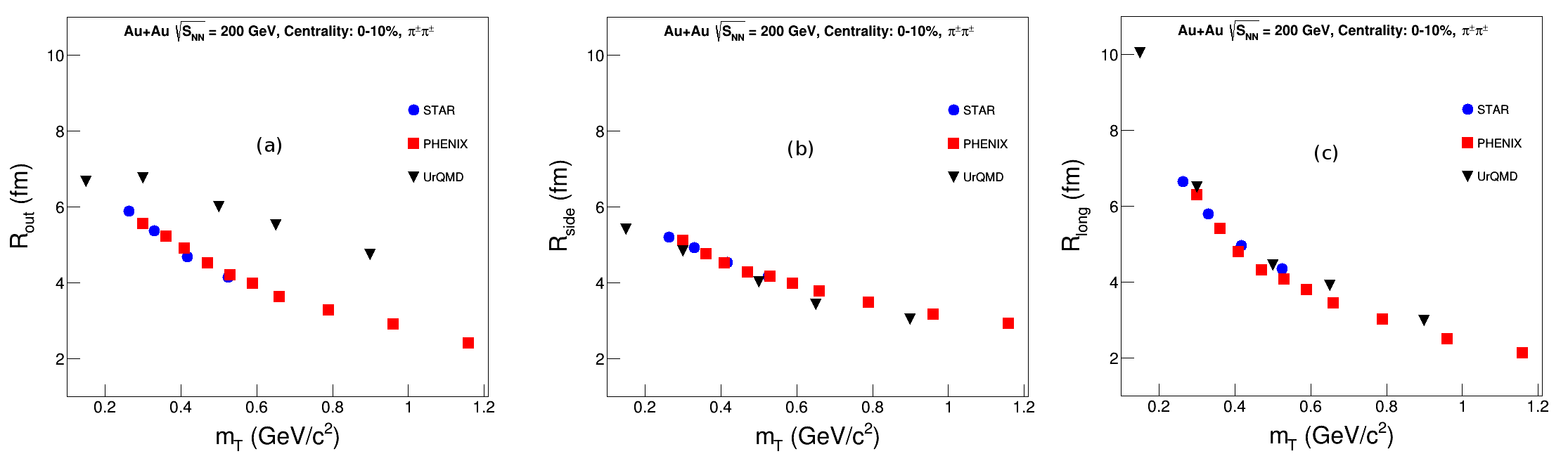}
  \caption{$R_{out}$ (a), $R_{side}$ (b) and $R_{long}$ (c) radii
    extracted from UrQMD model in comparison with STAR and PHENIX experiments.}
  \label{fig2}
\end{figure}

Fig.~\ref{fig2} shows the comparison of the $m_{T}$ dependence of the
HBT radii obtained from UrQMD model (black triangles) and the data
from STAR (blue circles) and PHENIX (red squares) experiments.

The $m_{T}$ dependence of $R_{long}$ and $R_{side}$ calculated
from UrQMD qualitatively and quantitatively agrees with the
experimental data. This suggests that the geometrical size of the
paricle emitting region and the medium lifetime are well reproduced by
the model.
From Fig.~\ref{fig2} it is seen that the $R_{out}$
obtained from the model is overestimated but qualitatively agrees
with the experimental data. Since the $R_{out}/R_{side}$ ratio
provides the information about the particle emission duration, one may
conclude that emission duration of the charged pions not well
reproduced in UrQMD.

\begin{figure}[h]
  \centering
  \includegraphics[width=0.8\textwidth] {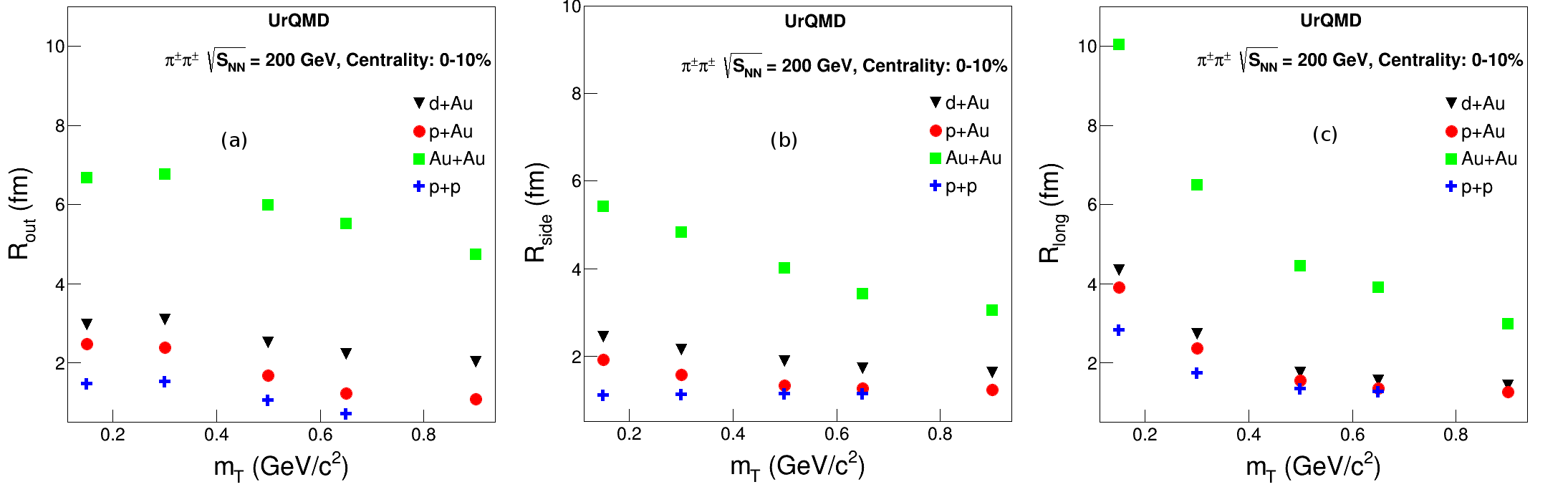}
  \caption{The $m_T$ dependences of the $R_{out}$ (a), $R_{side}$ (b)
    and $R_{long}$ (c) obtained for p+p, p+Au, d+Au and Au+Au.}
  \label{fig3}
\end{figure}

Fig.~\ref{fig3} shows the $m_T$ dependence of the charged pion
femtoscopic radii for p+p (blue crosses), p+Au (red circles), d+Au
(black triangles) and Au+Au (green squares) collisions obtained from
the UrQMD model.

It is seen that for each transverse mass range the source radii
following the next ordering:
$R_{i}$(Au+Au)$>R_{i}$(d+Au)$>R_{i}$(p+Au)$\geq R_{i}$(p+p). 
This estimations may be used for the future measurements at
Relativistic Heavy Ion Collider (RHIC) and for the correction of the
transport properties of the UrQMD model.

\section{Conclusions}

The two-pion femtoscopic correlations for p+p and central p+Au, d+Au
and Au+Au at top RHIC energy are studied with the Ultra Relativistic
Quantum Molecular Dynamics Model. The transverse mass dependence
of the femtoscopic radii $R_{out}, R_{side}, R_{long}$ were
extracted. For Au+Au collisions the estimated from the model radii
were compared to the experimental data from STAR and PHENIX.

The $m_{T}$ dependence of $R_{long}$ and $R_{side}$ radii obtained
from UrQMD qualitatively and quantitatively agrees with the
experimental data. The $R_{out}$ estimated by UrQMD are overestimated
which means that the emission duration of charged pions is
overestimated in the model meanwhile the geometrical size of the
particle emitting region and the medium lifetime are well reproduced.

The estimations of the $m_T$ dependence of the charged pion radii have
been done for p+p and central (0-10\% centrality) p+Au, d+Au,
Au+Au collisions. It was shown that for the given tranverese mass
region $R_{i}(Au+Au)>R_{i}(d+Au)>R_{i}(p+Au)\geq R_{i}(p+p)$.

\section*{Acknowledgments}
The reported study was funded by RFBR according to the research
project No. 16-02-01119 a.
Authors also would like to thank for the support from National
Research Nuclear University
MEPhI in the framework of the Russian Academic Excellence Project
(constract No. 02.a03.21.0005, 27.08.2013).
\section*{References}
\bibliographystyle{iopart-num}
\bibliography{Ermakov_UrQMD}
\end{document}